\documentclass{aa}
\input epsf.sty 
\begin{document}
\thesaurus{02.13.2;08.16.6;09.10.1}
\title{On the physics of cold MHD  winds from  oblique rotators.}
\subtitle{}
\author{S.V.Bogovalov}
\institute{Moscow Engineering Physics Institute, Kashirskoje Shosse 31,
Moscow, 115409, Russia. e-mail: bogoval@photon.mephi.ru}
\date{Received date today; accepted date yesterday}
\maketitle
\begin{abstract}
I show that the self-consistent solution of the problem of MHD plasma flow in 
 magnetosphere of an 
oblique rotator with an initially split-monopole magnetic field is reduced to the solution of 
the similar problem for the  axisymmetric rotator.
All properties of the MHD cold plasma flows    
from  the axisymmetric  rotators with the initially split-monopole
magnetic field are valid for the oblique rotators as well. 
Rotational
losses of the oblique rotator do not depend on the inclination angle and there is no
temporal evolution of this angle. 
Self - consistent analytical and numerical solutions for the axisymmetric plasma flows 
obtained earlier show that the rotators
can be divided on fast rotators ( $\sigma_0/U_{0}^{2} > 1$) and slow rotators 
( $\sigma_0/U_{0}^{2} < 1$),  where $\sigma_0$ is 
the ratio of
the Poynting flux  to the matter energy flux in the flow 
at  the equator on the surface of the star, 
$U_0=\gamma_{0}v_0/c$, $v_0$ and $\gamma_0$ being  the initial velocity and the 
Lorentz-factor of the plasma. The self-consistent approximate analytical 
solution for the plasma 
flow from the oblique  rotator  is obtained under the  condition $\sigma_0/U_{0}^{2} \ll 1$.
Implications of these results for   radio pulsars are discussed.
In particular, I argue that all radio pulsars are apparently the slow rotators ejecting
the Poynting dominated relativistic wind.

\end{abstract}
\section{Introduction}
An analysis of the
relativistic plasma flow is necessary for understanding  the processes
taking place  in radio pulsar's magnetospheres,
compact galactic objects and in AGN's 
(Arons \cite{arons96}; Mirabel \cite{mirabel}; Pelletier et al.
\cite{pelletier}). In the present paper,
as in previous one (Bogovalov \cite{bog97}), we
concentrate  our attention totally on the problem of the
relativistic plasma flow in the conditions typical for radio pulsars.

In spite of  systematic
research in the field of the physics of radio pulsars  in the last years,
the structure of the magnetosphere and the mechanisms for the acceleration
of plasma in these objects to large extent remain vague 
(Lyubarsky \cite{lyubarskii1995}).  One of the most important unsolved 
problem of the physics of radio pulsars is the problem of energetics of the 
ejected wind  of relativistic plasma. For example,  
the kinetic energy of the plasma accelerated in 
the inner magnetosphere of the Crab pulsar  is not sufficient 
to explain the energetics of the relativistic wind  exiting 
the synchrotron nebula surrounding the pulsar (Arons \cite{arons96}).

The basis of the theory of  the plasma production in  radio pulsar
magnetospheres was  initiated by Sturrock (\cite{sturrock}). This theory
was developed in more detail for different conditions on the stellar surface
 by Ruderman \& Sutherland (\cite{ruderman})
and by Arons (\cite{arons}). Primary electrons
are accelerated  in the so called
"electrostatic gaps" and produce
dense relativistic plasma  of secondary particles with  Lorenz-factor 
$\gamma_0 \sim 10^2-10^{3}$. This plasma screens the accelerating electric field and
 limits the potential drop by the value $\gamma_{gap}mc^2$ with 
$\gamma_{gap} \sim 10^7$. 
The total flux of the kinetic energy of all particles appears much less than the
total rotational losses
of the fast rotating radio pulsars due to this limitation. 
 Almost all energy is carried out by
the electromagnetic field. The wind with similar characteristics is formed in the
outer gap model (Cheng, Ho \& Ruderman \cite{cheng}).
The wind of  relativistic plasma from radio pulsars can be characterized by 
the ratio of the Poynting flux to
the flux of the
kinetic energy of the plasma  $\sigma$. The plasma is Poynting flux dominated when $\sigma > 1$
and is kinetic energy dominated if $\sigma < 1$.  Electrostatic gaps give  $\sigma > 1000$
for the Crab pulsar.
At the same time interpretation of observations of  the Crab Nebula compels us to
conclude 
that this ratio at large distances from the pulsar is very small.
This conclusion is based on the  assumption that the subsonic flow of the plasma
terminated by the shock wave at the interaction of the wind with the interstellar medium 
can be considered as the flow of ideal plasma with the only dissipative process of
synchrotron cooling. Under this assumption  the observed expansion of the outer edge of 
the Crab Nebula, the synchrotron  and  TeV gamma-ray emission
of the nebula produced via Inverse Compton Scattering of the relativistic particles on 
2.7-K MBR emission
can be explained under unique choice $\sigma = 3\cdot 10^{-3}$ 
(Atoyan \& Aharonian \cite{aharonian}).

Recently  Begelman (\cite{begelman}) has revisited  the key assumption of the
theory by Kennel \& Coroniti (\cite{kennel}). He argued that even a wind with 
$\sigma  \sim 1$
in the pre-shock region can give the observable properties of the Crab Nebula  
if one takes into
account dissipative processes in the nebula. 
It follows from the theory of relativistic MHD shocks that
a wind with $\sigma \sim 1$ in the pre-shock region creates the flow in the 
post-shock region with
$\sigma > 1$. But this flow  must be strongly unstable. The instability provides an effective
transformation of the magnetic field into the kinetic energy of the plasma 
accompanied by  acceleration of particles in the nebula.
It is important that in both  (Kennel \& Coroniti or Begelman) scenario it follows that  
there exists some unspecified mechanism for the transformation of the Poynting flux into 
the flux of the kinetic  energy as the plasma travels from  the star to  infinity.    
This mechanism is apparently the
basic mechanism for  the  plasma acceleration since it ensures the transformation
of at least 50\% ( in the Begelman scenario) or 99.7 \% ( in the Kennel \& Coroniti model)
of the rotational energy of the neutron star into the
kinetic energy of the relativistic wind. 

One of the possible mechanisms of the acceleration is the magnetic acceleration of the 
plasma 
by the  rotating magnetosphere. 
Unfortunately this mechanism appeared non effective for the axisymmetric 
rotators with typical pulsar's parameters.
No effective acceleration  was found neither in nearest zone ( Bogovalov \cite{bog97}), 
nor in far zone of the rotator (Bogovalov \& Tsinganos \cite{bogtsin}). 
At the same time up to now there is no solution
describing the MHD plasma flow from the oblique rotator. 
Usually it is believed that the inclination of the magnetic moment of the
star to the axis of rotation could be important for the plasma acceleration. In this 
paper we study this possibility in the model of the oblique rotator with the initially 
split-monopole
magnetic field on the surface of the star.

\section{Basic assumptions}

\begin{figure}
\epsfxsize=8.0 cm
\centerline{\epsfbox[0 0 290 270]{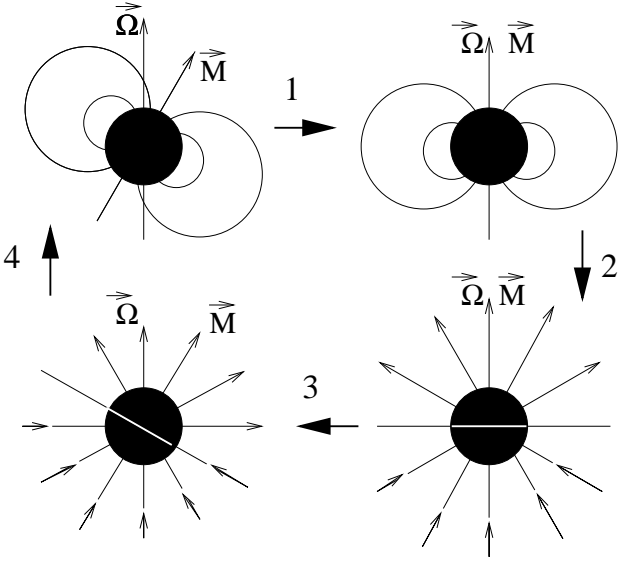}}
\caption{The sequence of models introduced to simplify the solution of the 
problem of the plasma outflows from radio pulsars. Step 1 is  the transition from the
oblique rotator with a dipole magnetic field to the axisymmetric rotator. Step 2 is the 
transition from the axisymmetric rotator with the dipole magnetic field to the axisymmetric 
rotator with the split-monopole magnetic field.  In this paper
the transition from the axisymmetric split-monopole to the oblique split-monopole 
model is done (step 3) together with step 4 connecting this model with real 
pulsars.}
\label{fig1}
\end{figure}

The axis of rotation of real radio pulsars is not directed along the
magnetic moment. The solution of the problem of the plasma flow in 
magnetosphere of this object is extremely complicated. To simplify the problem
several models of radio pulsars were proposed. 
The sequence of these models is presented in Fig. \ref{fig1}.
Firstly  Goldreich \& Julian (\cite{goldreich}) proposed  axisymmetrically
rotating star with initially dipole magnetic field
(step 1 in Fig. \ref{fig1}).
The rotational losses  of the axisymmetrically rotating star ejecting relativistic plasma 
are  comparable with the rotational losses of the oblique dipole in vacuum.
The energy of rotation is carried out near the surface
of the axisymmetric rotator by the Poynting flux. 
This is why this model can be used for study of 
the process  of the Poynting flux
transformation into the kinetic energy of the plasma. 

However even with such simplification the structure of the axisymmetric flow of the plasma 
from the star with the
initially dipole magnetic field 
appears too complicated. There were a lot of attempts to solve this problem in massless
approximation (Michel \cite{michel1973}; Beskin, Gurevich \& Istomin \cite{beskin}; 
Lyubarskii \cite{lubarskii1990}; Contopoulos,  Kazanas \& Fendt \cite{janis}). To simplify the problem  
Michel (\cite{michel69})
was the first who used
 the rotator with the prescribed split-monopole poloidal magnetic field
 for the investigation of the relativistic plasma flow in
the magnetosphere of the axisymmetrical rotator (step 2).
It follows from  Fig. \ref{fig1} that there are no closed field 
lines in the split-monopole model. All lines go to infinity.  This allows one
to remarkably simplify the analysis of the flow. 

Michel (\cite{michel69}) considered the flow of the plasma in the prescribed 
split-monopole magnetic field. The affect of the moving plasma on the poloidal
magnetic field was not taken into account. In the self-consistent solution the plasma 
and the electromagnetic field affect each other.    
The problem of the self-consistent plasma flow from the rotator with the  initially 
split-monopole magnetic field was investigated in nonrelativistic
and relativistic limits in the papers  by Sakurai (\cite{skurai}), 
Bogovalov (\cite{bog92}) Bogovalov (\cite{bog97}), Bogovalov \& Tsinganos (\cite{bogtsin}).
The phrase ``initially split-monopole magnetic field'' means that the 
normal component of the magnetic field on the surface of the star is constant but changes sign
on the magnetic equator. 
In this paper we firstly consider the model of the oblique rotator with the split-monopole
magnetic field on the surface of the star (step 3). 
This model allows one to investigate the plasma flow in conditions more typical
for real radio pulsars than it occurs in the axisymmetric model and to connect the
model with the split-monopole magnetic field with the real pulsars (step 4). 

Since the density of the relativistic plasma produced in the pulsar magnetosphere 
is high enough to screen the electric field parallel to the magnetic
field, the magnetohydrodynamical approximation is 
used in this paper to 
describe   the flow of the plasma.
The plasma is considered as cold  to a first approximation
(Lyubarsky \cite{lyubarskii1995}).

\section{Reduction of the oblique rotator problem to the axisymmetrical problem}

System of  time dependent equations defining the temporal evolution of the 
flow of the relativistic plasma  (Akhiezer et al \cite{akhiezer}) 
\begin{equation}
mn({\partial \gamma {\bf v}\over \partial t}+({\bf v\nabla})\gamma {\bf v})
=q\cdot {\bf E} +{1\over c} {\bf j}\times {\bf H},
\end{equation}
 \begin{equation}
{\partial {\bf H}\over c\partial t} = curl  {\bf E}, 
\end{equation}
 \begin{equation}
curl {\bf H} ={4\pi \over c}{\bf j} +{\partial {\bf E}\over c\partial t}, 
\end{equation}
\begin{equation}div{\bf H} =0,\end{equation}
\begin{equation}div{\bf E} =4\pi q,\end{equation}
\begin{equation}{\partial n\over \partial t}+ div ({\bf v}n) =0.\end{equation}
together with  frozen-in condition ${\bf E}+{1\over c} {\bf v}\times {\bf H}=0$ 
 gives  in spherical coordinate system the equations of motion
\begin{eqnarray}
\lefteqn{mn({\partial \gamma v_r\over \partial t}+({\bf v\nabla})\gamma v_r-
{\gamma(v_{\theta}^2+v_{\varphi}^2)\over r})=}\nonumber\\
&&q\cdot E_r +{1\over 4\pi}\Bigl\{({1\over r\sin{\theta}}{\partial H_r\over\partial\varphi}
-{1\over r}{\partial(rH_{\varphi})\over \partial r})H_{\varphi}-\nonumber\\
&& -({\partial(rH_{\theta})\over r \partial r}-{\partial H_r\over r \partial \theta})H_{\theta}+
{1\over c}(H_{\theta}{\partial E_{\varphi}\over \partial t}-H_{\varphi}{\partial E_{\theta}\over \partial t})\Bigr \},
\label{1}
\end{eqnarray}
\begin{eqnarray}
\lefteqn{mn({\partial \gamma v_{\theta}\over \partial t}+({\bf v\nabla})\gamma v_{\theta}+
{\gamma (v_{r}  v_{\theta}-v_{\varphi}^2 \cot\theta ) \over r})=}\nonumber\\
&&q\cdot E_{\theta} +{1\over 4\pi}\Bigl\{({\partial (rH_{\theta})\over r\partial r}
-{\partial H_{r}\over r\partial \theta})H_{r}-\nonumber\\
&& -{H_{\varphi}\over r\sin\theta}({\partial \sin\theta H_{\varphi}\over  \partial \theta}-
{\partial H_{\theta}\over \partial \varphi})+\nonumber\\
&&+{1\over c}(H_{\varphi}{\partial E_{r}\over \partial t}-H_{r}{\partial E_{\varphi}\over 
\partial t})\Bigr\},
\label{2}
\end{eqnarray}
\begin{eqnarray}
\lefteqn{mn({\partial \gamma v_{\varphi}\over \partial t}+({\bf v\nabla})\gamma v_{\varphi}+
{\gamma (v_r v_{\varphi}+v_{\theta}v_{\varphi}\cot\theta)\over r})=}\nonumber\\
&&q\cdot E_{\varphi} 
+{1\over 4\pi}\Bigl\{{1\over r\sin\theta}({\partial \sin\theta H_{\varphi}\over 
\partial \theta}-{\partial H_{\theta}\over \partial \varphi})H_{\theta}-\nonumber\\
&& -({1\over r\sin\theta}{\partial H_{r}\over  \partial \varphi}-
{\partial rH_{\varphi}\over r\partial r})H_{r}+\nonumber\\
&&+{1\over c}(H_{r}{\partial E_{\theta}\over \partial t}-H_{\theta}{\partial E_{r}\over 
\partial t})\Bigr\},
\label{3}
\end{eqnarray}

and  the induction equations
\begin{equation}
{\partial H_r\over c\partial t}=-{1\over r\sin\theta}({\partial (\sin\theta E_{\varphi})\over r \partial \theta}-{\partial E_{\theta}\over \partial\varphi}),
\label{i1}
\end{equation}
\begin{equation}
{\partial H_{\theta}\over c\partial t}=-({1\over r\sin\theta}{\partial E_{r}\over r \partial \varphi}-{\partial (rE_{\varphi})\over r\partial r}),
\label{i2}
\end{equation}
\begin{equation}
{\partial H_{\varphi}\over c\partial t}=-({\partial (rE_{\theta})\over r \partial r}
-{\partial E_{r}\over  r\partial\theta}).
\label{i3}
\end{equation}
The laws of conservation of the magnetic field  and  the matter flux,  
together with Coulomb law take a form
\begin{equation}
{1\over r^2}{\partial (r^2H_r)\over \partial r}+{1\over r\sin\theta}{\partial\sin\theta H_{\theta}\over \partial\theta}+
{1\over r\sin\theta}{\partial H_{\varphi}\over \partial\varphi}=0.
\label{m1}
\end{equation}
 \begin{equation}
{\partial n\over \partial t}+{1\over r^2}{\partial (r^2nv_r)\over \partial r}+{1\over r\sin\theta}{\partial\sin\theta nv_{\theta}\over \partial \theta}+
{1\over r\sin\theta}{\partial nv_{\varphi}\over \partial\varphi}=0.
\label{n1}
\end{equation}
\begin{equation}
{1\over r^2}{\partial (r^2E_r)\over \partial r}+{1\over r\sin\theta}{\partial\sin\theta E_{\theta}\over \partial \theta}+
{1\over r\sin\theta}{\partial E_{\varphi}\over \partial\varphi}=4\pi q.
\label{c1}
\end{equation}

The frozen-in conditions are
\begin{equation}
E_r+{1\over c} (v_{\theta}H_{\varphi}-v_{\varphi}H_{\theta})=0,
\label{f1}
\end{equation}
\begin{equation}
E_{\theta}+{1\over c} (v_{\varphi}H_{r}-v_{r}H_{\varphi})=0,
\label{f2}
\end{equation}
\begin{equation}
E_{\varphi}+{1\over c} (v_{r}H_{\theta}-v_{\theta}H_{r})=0.
\label{f3}
\end{equation}
In this system  $q$ is the induced space electric charge density, $\theta$ is the polar angle
and $\varphi$ is the azimuthal angle.

Boundary conditions on the surface of the star with radius $R_*$ for the system of 
equations above are:\\
1. The Lorenz-factor $\gamma_0$ is specified constant; \\
2. The normal component of the density of the matter flux is specified constant and 
uniform;\\
3. The normal component of the poloidal magnetic field $H_0$ does not depend on 
coordinates and time in every point of the  surface of the star and changes  sign
on the magnetic equator.\\
4. The tangential component of the electric field is continuous on the star surface; \\
It is assumed that  the velocity of the plasma   exceeds  fast magnetosonic
velocity (the flow is super sonic) at the infinity.
This system of equations together with the boundary conditions 
describes the stationary axisymmetric flow as well. 

\begin{figure}
\epsfxsize=8.0 cm
\centerline{\epsffile[0 0 588 250]{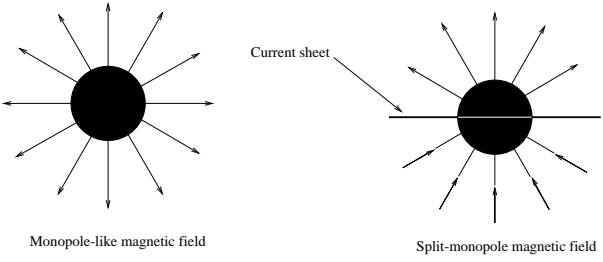}}
\caption{The solution in the model  with the split-monopole magnetic field can
be obtained from the solution for the monopole-like magnetic field by change of 
 sign  of the magnetic field in lower hemisphere. The same is valid for reverse operation.
There is  a current sheet at the equatorial plane in the split-monopole model which is a contact
discontinuity in ideal MHD approximation}   
\label{fig2}
\end{figure}

Let's assume that the self-consistent solution for  the problem of the plasma flow
from  the axisymmetric 
rotator with an  {\it initially monopole-like} poloidal magnetic field is specified.
The phrase  ``initially monopole-like magnetic field'' means that the normal component
of the poloidal magnetic field on the surface of the star is constant and does not 
change sign on the magnetic equator. Actually this field has no magnetic equator at all.
Although such magnetic fields can not be created in reality, this rather 
artificial mathematical model is convenient to construct solutions 
for  more realistic plasma flows. 
  The monopole-like  magnetic field differs from the split-monopole magnetic
field only by  direction of the field lines in one of hemispheres and was 
introduced by
Michel (\cite{michel69}). 
In the model with the monopole-like magnetic field all field lines
are out coming from the surface of the star as it is shown in Fig. \ref{fig2}.
There is no current sheet in this flow in contrast  to the model with the 
split-monopole magnetic field
which contains the current sheet in the equatorial plane. The current sheet is
a contact discontinuity (Landau \& Lifshitz \cite{lanel}) in ideal MHD approximation. 
 The magnetic field changes
 direction at the passage through the current sheet. It is obvious that  the solution
for the split-monopole can be obtained from the solution for the monopole-like magnetic filed
by formal reverse sign  of the magnetic fields in the lower hemisphere. Similar operation can be 
done in the case of the oblique rotator as well.

The flow and the 
monopole-like poloidal magnetic field are perturbed by rotation in the 
self-consistent axisymmetric flow (Bogovalov \& Tsinganos \cite{bogtsin}). We 
assume that this perturbed self-consistent solution is known.
Let's introduce the  velocity  
$  {\bf V(r)}$, the poloidal  $ {\bf B_p}({\bf r})$ and toroidal $B_{\varphi}({\bf r})$ 
magnetic fields and  the density of the plasma $N({\bf r})$ in the self-consistent 
axisymmetric flow  with the initially monopole-like magnetic field.
The poloidal electric
field ${\bf E}$ and the poloidal magnetic field are connected in the axisymmetric flow
${\bf E}=
(r \sin\theta\Omega / c)  {\bf B_p} \times \bf e_{\varphi}$
(Weber \& Davis \cite{weber}), where $\bf e_{\varphi}$ is the unit vector directed in the 
azimuthal direction. The toroidal electric field is equal to $0$. 
These variables depend only on coordinates. 
Dependence on time is absent for the stationary axisymmetric flow.
This solution
satisfies  automatically all  the equations of the system  (\ref{1}-\ref{f3}) and corresponding
boundary conditions. 

Now we show how the self-consistent solution for the plasma flow in the magnetosphere of
the  oblique rotator with the initially split - monopole magnetic field can be obtained
from  the known solution for the axisymmetric problem. 
Let's consider the following transformation
of the axisymmetric  solution. 
The velocity   and the density are taken  the same as for the axisymmetric rotator
\begin{equation}
v({\bf r},t)=V({\bf r}), ~~ n({\bf r},t)=N({\bf r}).
\label{s1}
\end{equation}
The poloidal magnetic field ${\bf H_p}({\bf r},t)$  for the oblique rotator is obtained
from the poloidal magnetic field of the axisymmetric rotator as follows
\begin{equation}
{\bf H_p}({\bf r},t)= \eta({\bf r},t){\bf B_p}({\bf r}) .
\label{s2}
\end{equation}
The same procedure gives us the toroidal magnetic field
\begin{equation}
H_{\varphi}({\bf r},t)= \eta({\bf r},t) B_{\varphi}({\bf r}),
\end{equation}
where $\eta({\bf r}, t)$ is unknown function to be specified.
The poloidal electric field $E$ is defined as in the axisymmetric case
\begin{equation}
E_{\theta}=-{r\sin\theta\Omega\over c} H_r,
\label{e1}
\end{equation}
\begin{equation}
E_{r}={r\sin\theta\Omega\over c} H_{\theta},
\label{e2}
\end{equation}
and
the toroidal electric field $E_{\varphi}=0$.
It is easy to show that this is  indeed  the solution of the problem for the oblique rotator
and to specify  the function $\eta$. 

The frozen-in conditions (\ref{f1}-\ref{f3})
are fulfilled automatically for the solution (\ref{s1}-\ref{e2}).
At the stationary rotation
the dependence of $\eta$ on $\varphi$ and $t$ can be presented by one variable
$\xi=\varphi-\Omega t$ in the spherical system of coordinates. Thus, $\eta$ depends on three
variables ${ r}$, $\theta$ and $\xi$. 
Then  the induction equation (\ref{i1}) is reduced to the
equation
\begin{equation}
{\partial\over \partial\xi}({E_{\theta}\over r\sin\theta}+ {\Omega H_r\over c})=0
\end{equation}
which  is fulfilled due to  condition (\ref{e1}). The same is valid for the
induction equation (\ref{i2}) which is reduced to the equation 
\begin{equation}
{\partial\over \partial\xi}({E_{r}\over r\sin\theta}- {\Omega H_{\theta}\over c})=0
\end{equation}
which  is satisfied due to  condition (\ref{e2}). Third   induction equation (\ref{i3})
after substitution of  conditions ({\ref{e1},\ref{e2}) is reduced to the 
flux conservation condition (\ref{m1}) which takes a form 
\begin{equation}
({\bf B}\cdot {\bf \nabla}\eta)=0.
\label{etaeq}
\end{equation}
since the total magnetic field  ${\bf B}$ of the axisymmetric flow 
satisfies  the equation ${\bf\nabla}\cdot {\bf B}=0$.
It implies that function $\eta$  is constant on the field line of the magnetic filed 
of the axisymmetric flow.
\begin{figure}
\epsfxsize=5.0 cm
\centerline{\epsffile[0 0 372 394]{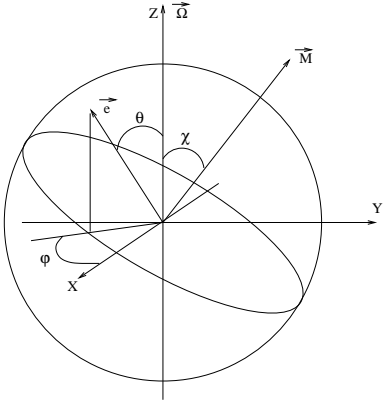}}
\caption{The initial position of the star and the 
geometry used in equation (\protect\ref{bceta}).}   
\label{fig3}
\end{figure}       

Fig. \ref{fig3}  shows the initial 
position   of the star  with the magnetic moment ${\bf M}$ inclined to the axis of rotation
at the angle $\chi$ at $t=0$.  It follows from 
equation (\ref{s2}) and the boundary conditions 
that the function $\eta$ is equal to 1 in the points on the surface of the star 
where the field line go out from the surface 
of the star and is equal to $-1$ in the points where the filed lines come into the surface. 
It is convenient to introduce function
\begin{equation} 
D(x) =\left\{
\begin{array}{rl}
1, & \mbox{if } x \ge  1\\
-1, & \mbox{if } x < 1.
\end{array} \right.
\end{equation}
The sign of the function $\eta$ on the surface of the star varies with the sign  of the product 
$({\bf e}\cdot {\bf e}_{M})$, where ${\bf e}$ is the unit vector directed to the point
on the surface of the star, ${\bf e}_{M}$ is the unit vector directed along the magnetic moment.  This product can be presented as
\begin{equation}
({\bf e}\cdot {\bf e}_{M})= \sin\chi\sin\theta\sin\varphi + 
\cos\chi\cos\theta.
\end{equation}  
Then, on the surface of the star, the function $\eta$ is
\begin{equation}
\eta(\theta,\varphi-\Omega t)=D(\sin\chi\sin\theta\sin(\varphi-\Omega t)+ 
\cos\chi\cos\theta).
\label{bceta}
\end{equation}
Actually this is the boundary condition to  equation (\ref{etaeq}).
The equation  can be presented in the form
\begin{equation}
B_r{\partial \eta\over \partial r}+ B_{\theta}{\partial \eta\over r\partial \theta}
+ B_{\varphi}{\partial\eta\over r\sin\theta\partial\varphi}=0.
\label{eta1}
\end{equation}
Equations of characteristics for this equation are
\begin{equation}
{dr\over B_r}={r \sin\theta d\varphi\over B_{\varphi}}
\end{equation}   
and 
\begin{equation}
{dr\over B_{r}}={r  d\theta\over B_{\theta}}.
\end{equation}
Therefore the general solution is
\begin{equation}
\eta(r, \theta, \varphi, t)= f(\theta-\int^r {B_{\theta}dr\over rB_r}, \varphi-
\int^r {B_{\varphi}dr\over r\sin\theta B_r}),
\end{equation}   
where the integrals over r are taken along the field line of the poloidal magnetic field of
the axisymmetric solution, $f$ is arbitrary function.
The solution  satisfying   boundary condition (\ref{bceta}) 
has a form
\begin{eqnarray}
\lefteqn{\eta(r, \theta, \varphi, t)=D\Bigl (\sin(\chi)\sin(\theta-\int^r_{R_*} 
{B_{\theta}dr\over rB_r})\times}\nonumber\\
&\times\sin(\varphi-\int^r_{R_*}  
{B_{\varphi}dr\over r\sin\theta B_r}-\Omega t)+\nonumber\\
&+\cos(\theta-\int^r_{R_*} {B_{\theta}dr\over rB_r})\cos\chi\Bigl ).&
\label{et}
\end{eqnarray} 

It follows from this solution  that $\eta^2=1$  
and $\eta$ changes sign when the magnetic field changes direction.
It is easy to show now that equations of
motion (\ref{1}-\ref{3}) are also satisfied for the solution (\ref{s1}-\ref{e2})
at the function $\eta$ defined by (\ref{et}). Notice that on the left hand side of these equations
there is no function $\eta$. In the right hand side of the equations of motion  function $\eta$ 
comes in combination $A_i \eta{\partial \eta B_k\over \partial x_l}$, where $A_i$ and $B_k$ are 
arbitrary components of fields of the axisymmetric solution, $x_l$ spatial or time 
coordinates in 4-space. This relationship can be presented as 
\begin{equation}  
A_i \eta{\partial \eta B_k\over \partial x_l}= \eta^2 A_i {\partial B_k\over \partial x_l} +
A_i B_k {1\over 2}{\partial \eta^2 \over \partial x_l}= A_i {\partial B_k\over \partial x_l}. 
\end{equation}
Therefore, the function $\eta$ disappears in the  equations of motion.  Here we ignore the 
difference in the dynamics of the current sheet and the surrounding plasma assuming that 
the current sheet is the mathematical discontinuity  as usual in ideal MHD. This 
assumption   can be violated for the oblique rotators at large distance from the star.
But at large distances the dynamics of the current sheet can be considered particularly 
in WKB approximation (Coroniti \cite{coroniti}; Michel \cite{michel94}).    

Thus, we obtain the self-consistent solution for the oblique rotator 
from the known self-consistent solution for the axisymmetric rotator.
The sketch demonstrating the structure of the cold wind from the oblique rotator is presented 
in Fig. \ref{fig4}. The structure of the plasma flow is symmetric in relation to the equator.
The form of the poloidal field lines is the same as for the axisymmetric rotator. In general
there is collimation of the plasma flow to the axis of rotation, although 
the effect of the collimation  depends on 
the parameters of the problem (Bogovalov \& Tsinganos \cite{bogtsin}). In the axisymmetric 
flow the current sheet dividing the magnetic fluxes of opposite directions is located
on the equator. In the wind from the oblique rotator   
the current sheet  takes a form of a wave. In the poloidal
plane the poloidal magnetic field lines change direction on the current sheet. 
At first glance it seems that this 
behavior  contradicts to the magnetic field freezing.  The bottom panel of  Fig.\ref{fig4} 
shows the structure of the field lines in the equatorial plane. 
It is seen that there is no
contradiction with the magnetic flux freezing since the total magnetic field 
depends on the azimuthal angle $\varphi$.   
\begin{figure}
\epsfxsize=8.0 cm
\centerline{\epsffile[0 0 704 901]{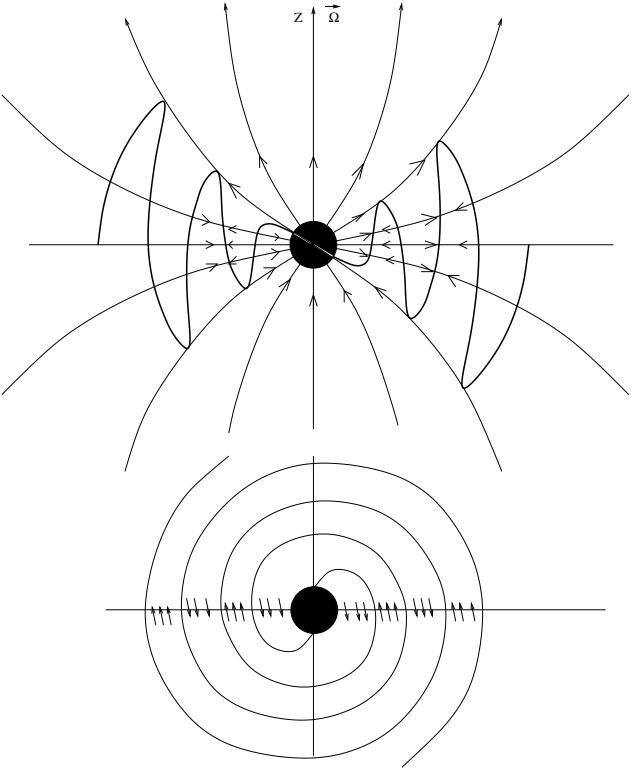}}
\caption{Top panel shows the structure of field lines and the current sheet (thick wave-like line)
in the poloidal plane. 
The lower panel shows the same in the equatorial plane. Arrows show the direction of the
magnetic field lines. The direction of the field lines changes on the current sheet.}   
\label{fig4}
\end{figure}       
The velocity and the density of the plasma do not depend on $\chi$ and $\varphi$ and 
are the same as for the
axisymmetric rotator. Only the magnetic and the electric fields are modulated with the
period of rotation. The sign of these fields is modulated in the sector limited
by the field lines which have roots on the magnetic equator. 
The squares of these fields are not modulated. The wave is simply the wave of the 
contact discontinuity (Landau \& Lifshitz \cite{lanel}) which is convected with 
the velocity of the plasma.   

The rotational losses of the oblique rotator can be calculated as the total 
Poynting flux through a sphere
surrounding the star.  The Poynting flux is defined by the formula 
$c{{\bf E\times H}\over 4\pi}$. This formula is biquadratic on the components
of the electromagnetic field. Since for the oblique rotator these components differ 
from the component for the axisymmetric rotator only on 
function $\eta$, the Poynting flux appears independent on the function $\eta$ and 
therefore the rotational losses are independent on the angle $\chi$. 

There is also no alignment of the magnetic moment along the axis  of
rotation as it happens in vacuum approximation
( Ostriker \& Gunn \cite{ostriker}, Pacini \cite{pachini}, 
Michel \& Goldwire \cite{michel70}, Davis \& Goldstein \cite{davis}) or disalignment
as it happens in the solution by Beskin, Gurevich \& Istomin (\cite{beskin}). 
The evolution of the 
inclination angle $\chi$ is defined by torques on the surface of the star
(Michel \& Goldwire \cite{michel70})
\begin{equation}
L_{y}=\int [T_{r\theta}\cos\varphi - T_{r\varphi}\cos\theta\sin\varphi]d^2 S,\nonumber
\end{equation}
\begin{equation}
L_{x}=-\int [T_{r\theta}\cos\varphi - T_{r\varphi}\cos\theta\sin\varphi]d^2 S,
\label{torque}
\end{equation}
and
\begin{equation}
L_{z}=\int T_{r\varphi}\sin\theta ~d^2S,\nonumber,
\end{equation}
where $T_{r\theta}$ and $T_{r\varphi}$ are the 
 components of the energy-momentum
tensor $T_{r\theta}$ and integration in (\ref{torque}) is performed over the surface of the 
star. 
The energy-momentum tensor is also biquadratic on the components of the electromagnetic field
(Landau \& Lifshitz \cite{landau}).
Therefore it does not depend on the function $\eta$ and the angles $\chi$ and $\varphi$.
The torques for the oblique rotator are the same as for the axisymmetric
rotator. Integration in eq. (\ref{torque}) over the surface of the star gives $L_x=L_y=0$.
There are no torques which can change the inclination angle of the oblique rotator
with the initially split-monopole magnetic field. This is not totally 
unexpected conclusion. The torques on the star rotating in
vacuum differs from the torque on the start ejecting ideal plasma.
The torque on the star rotating in vacuum always align the rotational and
magnetic axes (Soper \cite{soper}). The temporal evolution of the inclination angle $\chi$ of the star ejecting plasma depends on the distribution of 
the magnetic field on the surface of the star.  It was shown by perturbation
theory in first order approximation  on $\Omega$ that magnetic and rotation
axes tend to align when the magnetic flux is more concentrated to the magnetic poles and evolve to $\chi = 90^{\circ}$  when there is less flux density 
at the poles than at the equator (Mestel \& Selley \cite{mestel}). Thus,
the result obtained here for arbitrary angular velocity of the rotator
with the split-monopole magnetic field is due to combination of 
two conditions: by the ejection of the plasma simultaneously with 
the uniform magnetic flux distribution on the surface of the star.

\section{The flow of the plasma at ${\sigma_0\over U_0^2} \ll 1$.}

The solution of the problem of the cold  plasma flow from the rotator 
with the initially split-monopole magnetic field 
 is defined by the following parameters  and variables: $H_0$, $mn_0$,
$v_0$, $\Omega$, c, $r$, $\theta$, $\varphi$,  $R_{*}$,  $\chi$
and $t$. 
Lower index ``0'' denotes the values on the surface of the star. Since the dependence on 
$\chi$, $\varphi$ and $t$ is known, we can consider only the dependence 
of the solution for the axisymmetric
flow on other 8 parameters.  
There are 3 parameters in this set with independent dimension.
Therefore, according to the theory of dimensions (Barenblatt \cite{barenblat}) the solution in 
dimensionless variables depends only on 5 parameters. Let us write this dependence as
 \begin{equation}
{\bf A}={\bf f}(R_{star}/a, \alpha 1, \alpha 2,  r/a, \theta ).
\end{equation} 
Here ${\bf A}$ is the vector of the state of the plasma including density, velocity and magnetic
field, ${\bf f}$ is the unknown vector-function and $a$ is some parameter with dimension of 
length.  It is reasonable to consider limiting case 
when the dimensionless radius of the star goes to zero. 
It allows one to decrease  the number of variables
in the solution on one.  Now  the solution crucially depends only on two
parameters and two variables in the dimensionless presentation.

Let's measure the velocity of the plasma in units of the  initial 
velocity $v_0$ and 
all geometrical variables measure in the initial radius of the fast mode surface (FMS) $r_f$.
The FMS is the surface where the poloidal velocity of the plasma equals  the local velocity of
the fast mode MHD perturbations. 
In the comoving coordinate system where the electric field is equal 
to zero, this  condition takes a form 
\begin{equation}
v_p={H^{'2}\over 4\pi mn^{'}c^2+H^{'2}},
\end{equation}
where $n^{'}$ and $H^{'}$ are the density of the plasma and the magnetic filed in 
the comoving system. Taking into account that 
$H^2 -E^2$ is invariant (Landau \& Lifshitz \cite{landau}), $r_f$ can be calculated 
 at $\Omega =0$ as follows 
\begin{equation}
r_f^2={H_0^2R_{*}^2\over 4\pi mcn_0v_0U_0}.
\end{equation}
In these units the radius-vector is ${\bf X} = {\bf r}/r_f$.
The electric and magnetic fields can be measured in the units of the 
initial poloidal magnetic field
on the FMS. The density of the plasma can also be measured in units of 
the initial density on the 
FMS.

The parameters $\alpha 1$ and $\alpha 2$ can be chosen rather arbitrary. In particular it is
convenient to choose $\alpha 1 =\alpha= (r_f\Omega/v_0\gamma_0)$ and 
$\alpha 2 =\epsilon = (r_f\Omega/c)$. 
Let's consider the dependence of the solution on these parameters.

Nonrelativistic limit corresponds $\epsilon \rightarrow 0 $ and $v_0 \ll c$. The 
solution  depends only on one parameter in this limit.
The axisymmetrical nonrelativistic plasma  flow was investigated  
numerically  by Bogovalov \& Tsinganos (\cite{bogtsin}) 
for the parameter
$\alpha$ achieving the value up to 4.5. It was found that there is clear division
of the rotators on the fast  and  slow  in dependence on the value of $\alpha$. 
The slow rotators correspond to $\alpha < 1$
and the fast rotators correspond to $\alpha > 1$. 
There is strong physical difference between them. The plasma in outflows from
the slow rotators is not accelerated considerably and is not collimated in the subsonic region. 
In the case of the fast rotators the
plasma is effectively accelerated and collimated to the axis of rotation already in the 
subsonic region of the flow.

The flow of the relativistic plasma depends crucially on two parameters. But in this case
it is also possible to divide the rotators on
the slow and the fast ones depending on value of  the parameter $\alpha$. 
The solution can be presented as a vector function ${\bf A}$ with components $A_l$ which
are density of plasma, components of velocity, magnetic field and etc. 
Let's consider the expansion of this function
on  powers of $\alpha$ in the point $\alpha=0$ at fixed $\epsilon \ne 0$. Physically it means
that we consider the relativistic limit of the problem at $\gamma_0\rightarrow \infty$. The
expansion can be presented as  
 \begin{equation}
 A_l=f^{(0)}_l(\epsilon , {\bf X})(1+ \alpha {\bf f}^{(1)}(\epsilon, {\bf X})+
(\alpha)^2 {\bf f}^{(2)}(\epsilon, {\bf X})+ ....).
\label{range}
\end{equation} 
The first term $f^{(0)}_l(\epsilon , {\bf X})$ in this expansion has very simple presentation 
at arbitrary parameter $\epsilon \ne 0$. 
It was shown  that at the limit under consideration the function $f^{(0)}_l(\epsilon , {\bf X})$ 
is following  in the ordinary (dimension) variables
(Bogovalov, 1997)
\begin{equation}
U_r=U_0,  ~~U_{\varphi}=0,~~U_{\theta}=0,~~n(r)=n_0({r_0\over r})^2.
\label{sol1}
\end{equation}
\begin{equation}
H_{\varphi}=-({r\Omega\sin\theta\over c})H_p, ~~E=({r\Omega\sin\theta\over c})H_p,
\label{sol2}
\end{equation} 
and
\begin{equation}
H_r=\left\{\begin{array}{rl}
H_0({r_0\over r})^2, & \mbox{if } \theta \le  \pi/2\\
& \\
-H_0({r_0\over r})^2, & \mbox{if } \theta  >  \pi/2
\end{array} , ~~~H_{\theta}=0. \right.
\label{sol3}
\end{equation}
This is exactly the solution obtained by Michel (\cite{michel}) 
for the massless plasma.
It is easy to show that the full system of the equations of motion is satisfied for this solution
except the frozen-in condition (\ref{f3}). The residual in this condition  on the FMS is
\begin{equation}
\delta v_{\varphi}= \epsilon{(\alpha/ \epsilon)^2\over (1+v_0/c)}.
\end{equation}
The residual shows that the corrections to the solution are indeed proportional to
powers of $\alpha$ and can be neglected in the region limited by the FMS provided that
 $\alpha \ll 1$.
This residual also shows that expansion (\ref{range}) contains terms 
$(\alpha /\epsilon )^k$, where k is integer number. Therefore the first term in  
expansion (\ref{range}) gives incorrect  result in the limit $\alpha \rightarrow 0$, 
$\epsilon \rightarrow 0$  provided that
$\alpha/\epsilon = const$ which corresponds to the slow rotation of the star 
at arbitrary velocity $v_0$. Mathematically it
means that the point ($\alpha=0$, $\epsilon=0$) is a particular point of the expansion. 
The terms $(\epsilon /\alpha )^k$ can be eliminated from  expansion (\ref{range}) if to define  
the function
\begin{equation}
 g_l(\alpha/ \epsilon, {\bf X})=\lim\limits_{\begin{array}{c} \alpha \rightarrow 0\\ 
\alpha/\epsilon =const \end{array}} 
{A_l(\alpha, \epsilon, {\bf X})\over 
f^{(0)}_l(\epsilon , {\bf X})}.
\end{equation}
Then the expansion can be presented as
\begin{equation}
 A_l=f^{(0)}_l\bigl (\epsilon , {\bf X})(g(\alpha/\epsilon,{\bf X})
+\sum_{k=1} \alpha^k g_1^{(k)}(\epsilon, {\bf X}) \bigr ).
\end{equation}
The expansion $ \sum_{k=1} \alpha^k g_1^{(k)}(\epsilon, {\bf X})$ already does not contain
terms $(\alpha/\epsilon)^k$ and  function 
$ F_l=f^{(0)}_l(\epsilon , {\bf X})(g(\alpha/\epsilon,{\bf X})$ gives correct first term 
at $\alpha\rightarrow 0$ for arbitrary parameter $\epsilon$ including point $\epsilon=0$.
This function  was 
calculated analytically by Bogovalov (1992). The corrected solution of the problem in the 
limit $\alpha\rightarrow 0$  at arbitrary $\epsilon$ coincides with 
the solution (\ref{sol1}-\ref{sol3}) 
with exception of the  toroidal magnetic field. It is replaced by
\begin{equation}
H_{\varphi}= -({r\Omega\sin\theta \over v_0}) H_r.
\label{sol4}
\end{equation}

According to  this solution the plasma moves radially in the 
split-monopole poloidal magnetic field with
constant poloidal velocity and without any toroidal motion. At large distances 
the collimation of the plasma to the axis of rotation 
and corresponding acceleration of the plasma are also very weak 
(Bogovalov \& Tsinganos \cite{bogtsin}). It is reasonable to call  all rotators 
with $\alpha < 1$ as the slow rotators and   the rotators
with $\alpha > 1$ as the fast rotators independently on the value of $\epsilon$.

Let's consider the physical sense of the parameters $\alpha$ and $\epsilon$.
It is clear that $\epsilon$ is the ratio of the initial radius of the fast mode surface
to the light cylinder. On the other hand 
\begin{equation}
\epsilon^2= \sigma_0={H_0^2\over 4\pi n_0mc^2\gamma_0}({R_{*}\Omega\over v_0})^2.
\end{equation} 
For the slow rotators $\sigma_0$ is the ratio of the Poynting flux to the matter
energy flux (kinetic energy flux for the relativistic plasma) 
at the equator and coincides with the well known magnetization 
parameter (Arons \cite{arons96}) for the slow rotators. It is easy to show that 
\begin{equation}
\alpha^2={\sigma_0\over U_0^2}.
\end{equation}
Therefore the slow rotators correspond to the condition ${\sigma_0\over U_0^2} \ll 1$.

\begin{figure}
\epsfxsize=8.0 cm
\centerline{\epsffile[0 0 608 618]{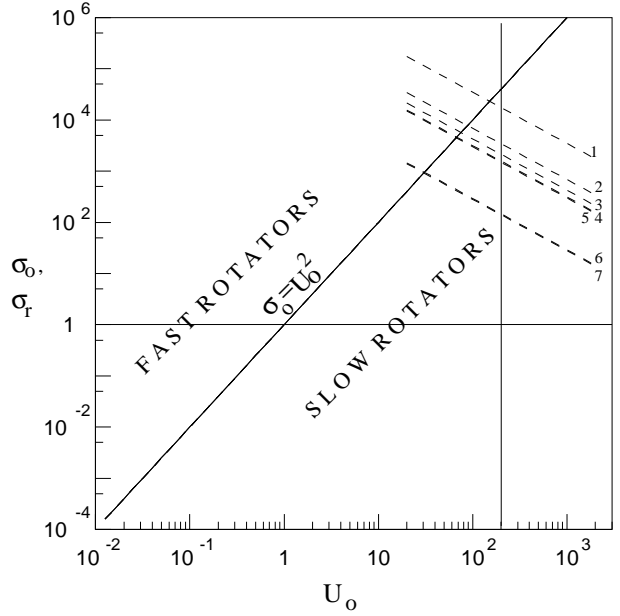}}
\caption{Regimes of the cold plasma flows.
The locations of 6 pulsars emitting observable gamma-rays above 100 MeV: Crab -1,
PSR 1508-59 - 2, Vela - 3,  PSR 1706-44 - 4, PSR 1951+32 - 5, Geminga -6 and PSR 1055-52 - 7 are
shown by dashed lines for all possible values of $U_0$.  
All the pulsars are located in the region of slow rotators ejecting the Poynting flux dominated
wind provided that $U_0 > 200$.}
\label{fig5}
\end{figure}       
Diagram in Fig. \ref{fig5} shows different regimes of the flow from rotators
with the initially split-monopole magnetic field in coordinates $\sigma_0$ and $U_0$.
The line $\sigma_0=1$ divides the rotators with the Poynting flux dominated flow 
near the surface of the star (above this line) from the rotators 
with  the matter energy dominated flow (below this line). 
The slow rotators are located below the line $\sigma_0= U_0^2$ while the fast rotators
are located above this line.
 In wide range of the parameters the regimes of the axisymmetric
flows have been already investigated. Now it is evident that the obtained solutions are 
valid for the oblique rotators also. 
For the slow rotators with the strongly matter energy dominated flow ($\sigma_0 \rightarrow 0$)
an approximate analytical solution with accuracy to the terms of the order $\alpha^2$ 
was obtained  by Bogovalov (\cite{bog92}). 
At the increase of $\sigma_0$ at constant $U_0$, 
rotators with the nonrelativistic flow firstly cross the line
$\sigma_0 =U_0^2$ and becomes fast rotators. Numerical solutions for the nonrelativistic 
plasma flows  ($\sigma_0 \rightarrow 0$) from fast rotators were obtained up to $\alpha=4.5$
(Bogovalov \& Tsinganos \cite{bogtsin}). It was found that due to the strong collimation of
the plasma to the axis of rotation, the flow  becomes subsonic   in some
region along the axis of rotation and therefore should be absolutely  
unstable well before the rotator  
reaches the  line $\sigma_0=1$ (see also Lyubarskii (\cite{lub}) and Begelman (\cite{begelman})). 

The rotators  with the initially relativistic plasma flow firstly cross the line $\sigma_0=1$
at the increase of $\sigma_0$ at constant $U_0$
and became the slow rotators ejecting the Poynting flux dominated wind.
At the increase of $\sigma$ they also 
cross the line $\sigma = U_0^2$ and become the fast rotators. 
The relativistic
stationary  plasma flow from the fast rotators 
was investigated 
recently by Beskin, Kuznetsova \& Rafikov (\cite{vasia98}). Surprisingly, the 
authors found that the 
perturbation of the poloidal  magnetic field go to zero as $\sigma_0/U^2_0 \rightarrow 0$
and acceleration of the plasma remains very non effective like in the  model by
Michel (\cite{michel69}) with the prescribed monopole-like poloidal magnetic field. 
The Lorentz-factor of the plasma reaches the value 
$\gamma_{\infty} \sim (\sigma_0\gamma_0)^{1/3}$.

The plasma flow from the slow oblique rotator becomes especially
simple. 
In particular, the function $\eta$  for the slow rotators ($\alpha \ll 1$) is
\begin{eqnarray}
\lefteqn{\eta(r,\theta,\varphi,t) = D(\sin\chi\sin\theta\sin(\varphi-\Omega(t-r/v_0))+}\nonumber\\
&&+ \cos\theta\cos\chi).
\end{eqnarray}

Rotational losses of the slow rotators with the initially split-monopole magnetic field 
are defined  by the total Poynting flux through the surface of the star 
\begin{equation}
\dot E_{rot} = {2\over 3}{H_0^2R_{*}^4\Omega^2\over v_0},
\end{equation}
and they do not depend on the angle $\chi$. It is useful also to 
express the rotational losses through the total magnetic flux $\psi$ of the 
open field lines of one direction (for example outcoming from the star)
\begin{equation}
\dot E_{rot} = {1\over 6\pi^2}{\psi^2\Omega^2\over v_0},
\end{equation} 
where $\psi=2\pi H_0 R_*^2$ for the split-monopole magnetic field.

\section{Implications to radio pulsars.}

How the model considered above is connected with the real radio pulsars?
It is reasonable to assume that the rotator with the split-monopole 
magnetic field  is equivalent to real radio pulsar if it provides
the plasma flow similar to the plasma flow from the real pulsar
at distances larger than the light cylinder.  
Therefore the equivalent split-monopole rotator  should have the 
matter flux, the poloidal magnetic field flux from one of the poles, 
$\gamma_0$, $v_0$, inclination 
angle $\chi$ and $\Omega$ the same as the real pulsar has.
 
The magnetic flux from one of the poles of the split-monopole rotator
is $\psi=2\pi R_{*}^2 H_0$. The magnetic flux of the open field lines
from the radio pulsar with dipole
magnetic field is defined by the radius of the polar cap 
$R_p=R_{*}\sqrt{{R_{*}\Omega\over c}}$ (Goldreich \& Julian \cite{goldreich}).
Therefore the flux of the open field lines from one of the polar caps of
the radio pulsar is estimated as
\begin{equation}
\psi = \pi H_0 R^2_{*}({R_{*}\Omega\over c}).
\end{equation}
The ratio of the poloidal magnetic field to the matter flux density $H_p/ nv_p$ is 
constant on the field line. Therefore this ratio for the  split-monopole rotator and for
the radio pulsar should be the same.  These conditions allow one to define all  
parameters. The split-monopole rotator equivalent to the pulsar with the magnetic field
on the polar cap $H_0$, angular velocity $\Omega$ and ejecting plasma with initial 
density $n_0$ and $\gamma_0$ has the magnetization parameter 
\begin{equation}
\sigma_r ={H_0^2\over 8\pi n_0mc^2\gamma_0}({R_{*}\Omega\over c})^3.
\label{sr}
\end{equation}  
 Below we consider the relativistic plasma with $v_0 =c$.
The rotational losses of this object are
\begin{equation}
\dot E_{rot} = {1\over 6}{H_0^2R_{*}^6\Omega^4\over c^3}
\label{lossesr}
\end{equation} 
These rotational losses equal  the losses of the  dipole 
rotating in vacuum at the angle of inclination $\chi = 30^0$.

The initial density of the plasma $n_0$ is defined by the processes of multiplication
of plasma in the electromagnetic cascades initiated by primary
particles with the Lorentz-factor $\gamma_{gap}\sim 2\cdot 10^7$ in the magnetosphere of pulsar.
In the inner gap theories (Ruderman \& Sutherland \cite{ruderman}; Arons \cite{arons})
the primary beam has Goldreich-Julian density 
\begin{equation}
n_{GJ} \sim {\Omega H_0\over 2\pi ec},
\label{ngj}
\end{equation}
We neglect here the dependence of $n_{GJ}$ on the inclination angle.
The primary particles produce $\lambda$ secondary electrons and positrons.
Therefore the initial density of the plasma is $n_0=\lambda n_{GJ}$.
Calculations of cascades performed by Daugherty \& Harding (\cite{daugherty}) and
by Gurevich \& Istomin (\cite{gurevich}) show that $\lambda \sim 10^4$.
Outer gap model (Cheng \& Ruderman \cite{cheng}; Romani \cite{romani})
gives similar values of $n_0$.
After substitution of (\ref{ngj}) in (\ref{sr}),  $\sigma_r$  takes a form
\begin{equation}
\sigma_r= {eH_0\over 4\lambda mc \Omega\gamma_0}({R_{*}\Omega\over c})^3.
\end{equation}

The estimates of $\gamma_0$ are less definite. The flow of the plasma formed in the
electromagnetic cascade consists of two components. One of them is the beam of the primary
particles which can loss remarkable amount of energy.  Another component is the plasma of 
the secondary  particles. 
It is reasonable to take the average Lorentz-factor of the secondary 
particles as $\gamma_0$ because it is this plasma  which screens the electric field and 
ensures the frozen-in condition. The average Lorentz-factor of the secondary particles 
strongly depends on the model of the gap. This parameter lies in the range  
$20 < \gamma_0 < \gamma_{gap}/\lambda$. 
To be definite we assume here that 
\begin{equation}
\gamma_0=0.1{\gamma_{gap}\over \lambda} \sim 200.
\end{equation} 
This value well agrees with the results of cascade simulations performed 
by Daugherty \& Harding (\cite{daugherty}). For other $\gamma_0$ the magnetization
parameter is scaled as $\sigma_r= \sigma_{200}(200/\gamma_0)$.
The location of all radio pulsar observed by EGRET in gamma-rays above 100 MeV 
(Crab, Vela, Geminga, PSR 1706-44,
PSR 1509-58, PSR 1951+32 and PSR 1055-55; Thompson \cite{thompson}) 
in the parameter space $\sigma_r$, $U_0$  for all possible $U_0$ 
is shown  in Fig. \ref{fig5} by dashed lines. 
For the estimates of the magnetic field $H_0$ equation (\ref{lossesr}) was used.
It was assumed also that the momentum of inertia of all pulsars is $10^{45} {\rm g\cdot cm^2}$.
It is seen that all  the  pulsars appears the slow rotators provided that 
$U_0 \ge 200$. This implies that they 
do not accelerate plasma and do not collimate it considerably  to the axis of rotation
beyond the light cylinder if the flow of the plasma is dissipativeless 
(Bogovalov \cite{bog98}; Bogovalov \& Tsinganos \cite{bogtsin}).

\section{Conclusion}

This paper clearly demonstrates that the oblique rotators  do not 
differ strongly from the axisymmetric rotators. In the particular case considered 
in this paper the dynamics of the plasma ejected by the oblique rotator is exactly 
the same as the plasma dynamics of the wind from the axisymmetric rotator. It 
allows us to apply the all results obtained for the cold axisymmetric plasma flows  
to the oblique rotators. But these results show that for  parameters typical for
radio pulsars the acceleration of the relativistic plasma is not effective. 
Radio pulsars apparently are basically slow rotators. The obliqueness of the rotators 
does not change this conclusion.  

Our results indicate that only violation of the ideal MHD can provide 
effective acceleration of the winds from radio pulsars  
since neither magneto-centrifugal mechanism or
the acceleration by the gradient of the toroidal magnetic field at the
collimation of the plasma to the axis of rotation are effective for the pulsars. 
The role of the dissipative processes in the relativistic winds 
is under discussion for  many years (Coroniti \cite{coroniti}, Michel \cite{michel94},
Mestel \& Shibata \cite{shibata}).
Recently  Melatos \& Melrose (\cite{melatos}) tried to argue that the 
violation of the frozen-in condition takes place in the wind from an oblique rotator
and it can provide the
acceleration of the wind. 
Our self-consistent solution of the problem for the particular case of the 
oblique rotator with
the initially split-monopole magnetic field shows no evidence of the violation of the ideal MHD in the wind. 

Mestel \& Shibata (\cite{shibata})  argued that the frozen-in
condition can be violated 
right after the light cylinder due to the formation of singularities in 
the stationary axisymmetric flow. This assumption is not confirmed by last 
results  by Bogovalov (\cite{bog97}) and 
by Contopoulos, Kazanas \& Fendt (\cite{janis}). 

It follows  from all results obtained recently in the physics of relativistic
plasma outflows that only dissipative processes can ensure plasma acceleration
under pulsar conditions. One of these mechanisms was proposed by 
Coroniti (\cite{coroniti}) and Michel (\cite{michel94}). They 
argued that  due to the geometrical reasons the dissipative 
processes can ensure  the 
transformation of the Poynting flux into the kinetic energy of the 
plasma provided that the inclination angle is high enough ($\sim 80^{\circ}$).
The Poynting flux is transformed into  the thermal energy of the plasma which
is than accelerated by the thermal pressure gradient.

Another possibility is the Poynting flux transformation in the volume of the flow.  
This can happen if the Poynting flux dominated plasma flow is unstable in relation
to MHD perturbations. The anomalously low conductivity due to the 
MHD turbulence would
provide the violation of the ideality in the flow  with corresponding 
transformation of the Poynting flux into the energy of the plasma. The advantage
of this approach is that it can provide effective acceleration even at the
axisymmetric rotation. But the 
question about stability of the Poynting flux dominated plasma flow 
is still open and should be investigated in details separatly.

\begin{acknowledgements} 
Author acknowledge discussions with colleagues from Astrophysics group of the MPI f\"ur Kernphysik  
and Dr. Yu.E.Lyubarsky. Author thanks Dr. F.A. Aharonian for reading of the 
manuscript and useful comments as well as 
MPI f\"ur Kernphysik for warm hospitality and support during the  work on the paper.
Author is also grateful to unknown referee for useful remarks on the paper.
\end{acknowledgements}

\end{document}